\documentclass[aps,prl,twocolumn,showpacs,superscriptaddress,letterpaper,amsmath,amssymb]{revtex4}  
\usepackage{graphicx} 
\usepackage{dcolumn}  
\usepackage{bm}       
\usepackage{amssymb}  
\usepackage{amsmath}

\begin{document}
\bibliographystyle{apsrev}

\title{A Search for the Higgs Boson Produced in Association with $Z\to \ell^+\ell^-$ Using the Matrix Element Method at CDF II}

\affiliation{Institute of Physics, Academia Sinica, Taipei, Taiwan 11529, Republic of China} 
\affiliation{Argonne National Laboratory, Argonne, Illinois 60439} 
\affiliation{University of Athens, 157 71 Athens, Greece} 
\affiliation{Institut de Fisica d'Altes Energies, Universitat Autonoma de Barcelona, E-08193, Bellaterra (Barcelona), Spain} 
\affiliation{Baylor University, Waco, Texas  76798} 
\affiliation{Istituto Nazionale di Fisica Nucleare Bologna, $^y$University of Bologna, I-40127 Bologna, Italy} 
\affiliation{Brandeis University, Waltham, Massachusetts 02254} 
\affiliation{University of California, Davis, Davis, California  95616} 
\affiliation{University of California, Los Angeles, Los Angeles, California  90024} 
\affiliation{University of California, San Diego, La Jolla, California  92093} 
\affiliation{University of California, Santa Barbara, Santa Barbara, California 93106} 
\affiliation{Instituto de Fisica de Cantabria, CSIC-University of Cantabria, 39005 Santander, Spain} 
\affiliation{Carnegie Mellon University, Pittsburgh, PA  15213} 
\affiliation{Enrico Fermi Institute, University of Chicago, Chicago, Illinois 60637}
\affiliation{Comenius University, 842 48 Bratislava, Slovakia; Institute of Experimental Physics, 040 01 Kosice, Slovakia} 
\affiliation{Joint Institute for Nuclear Research, RU-141980 Dubna, Russia} 
\affiliation{Duke University, Durham, North Carolina  27708} 
\affiliation{Fermi National Accelerator Laboratory, Batavia, Illinois 60510} 
\affiliation{University of Florida, Gainesville, Florida  32611} 
\affiliation{Laboratori Nazionali di Frascati, Istituto Nazionale di Fisica Nucleare, I-00044 Frascati, Italy} 
\affiliation{University of Geneva, CH-1211 Geneva 4, Switzerland} 
\affiliation{Glasgow University, Glasgow G12 8QQ, United Kingdom} 
\affiliation{Harvard University, Cambridge, Massachusetts 02138} 
\affiliation{Division of High Energy Physics, Department of Physics, University of Helsinki and Helsinki Institute of Physics, FIN-00014, Helsinki, Finland} 
\affiliation{University of Illinois, Urbana, Illinois 61801} 
\affiliation{The Johns Hopkins University, Baltimore, Maryland 21218} 
\affiliation{Institut f\"{u}r Experimentelle Kernphysik, Universit\"{a}t Karlsruhe, 76128 Karlsruhe, Germany} 
\affiliation{Center for High Energy Physics: Kyungpook National University, Daegu 702-701, Korea; Seoul National University, Seoul 151-742, Korea; Sungkyunkwan University, Suwon 440-746, Korea; Korea Institute of Science and Technology Information, Daejeon, 305-806, Korea; Chonnam National University, Gwangju, 500-757, Korea} 
\affiliation{Ernest Orlando Lawrence Berkeley National Laboratory, Berkeley, California 94720} 
\affiliation{University of Liverpool, Liverpool L69 7ZE, United Kingdom} 
\affiliation{University College London, London WC1E 6BT, United Kingdom} 
\affiliation{Centro de Investigaciones Energeticas Medioambientales y Tecnologicas, E-28040 Madrid, Spain} 
\affiliation{Massachusetts Institute of Technology, Cambridge, Massachusetts  02139} 
\affiliation{Institute of Particle Physics: McGill University, Montr\'{e}al, Qu\'{e}bec, Canada H3A~2T8; Simon Fraser University, Burnaby, British Columbia, Canada V5A~1S6; University of Toronto, Toronto, Ontario, Canada M5S~1A7; and TRIUMF, Vancouver, British Columbia, Canada V6T~2A3} 
\affiliation{University of Michigan, Ann Arbor, Michigan 48109} 
\affiliation{Michigan State University, East Lansing, Michigan  48824}
\affiliation{Institution for Theoretical and Experimental Physics, ITEP, Moscow 117259, Russia} 
\affiliation{University of New Mexico, Albuquerque, New Mexico 87131} 
\affiliation{Northwestern University, Evanston, Illinois  60208} 
\affiliation{The Ohio State University, Columbus, Ohio  43210} 
\affiliation{Okayama University, Okayama 700-8530, Japan} 
\affiliation{Osaka City University, Osaka 588, Japan} 
\affiliation{University of Oxford, Oxford OX1 3RH, United Kingdom} 
\affiliation{Istituto Nazionale di Fisica Nucleare, Sezione di Padova-Trento, $^z$University of Padova, I-35131 Padova, Italy} 
\affiliation{LPNHE, Universite Pierre et Marie Curie/IN2P3-CNRS, UMR7585, Paris, F-75252 France} 
\affiliation{University of Pennsylvania, Philadelphia, Pennsylvania 19104}
\affiliation{Istituto Nazionale di Fisica Nucleare Pisa, $^{aa}$University of Pisa, $^{bb}$University of Siena and $^{cc}$Scuola Normale Superiore, I-56127 Pisa, Italy} 
\affiliation{University of Pittsburgh, Pittsburgh, Pennsylvania 15260} 
\affiliation{Purdue University, West Lafayette, Indiana 47907} 
\affiliation{University of Rochester, Rochester, New York 14627} 
\affiliation{The Rockefeller University, New York, New York 10021} 
\affiliation{Istituto Nazionale di Fisica Nucleare, Sezione di Roma 1, $^{dd}$Sapienza Universit\`{a} di Roma, I-00185 Roma, Italy} 

\affiliation{Rutgers University, Piscataway, New Jersey 08855} 
\affiliation{Texas A\&M University, College Station, Texas 77843} 
\affiliation{Istituto Nazionale di Fisica Nucleare Trieste/Udine, I-34100 Trieste, $^{ee}$University of Trieste/Udine, I-33100 Udine, Italy} 
\affiliation{University of Tsukuba, Tsukuba, Ibaraki 305, Japan} 
\affiliation{Tufts University, Medford, Massachusetts 02155} 
\affiliation{Waseda University, Tokyo 169, Japan} 
\affiliation{Wayne State University, Detroit, Michigan  48201} 
\affiliation{University of Wisconsin, Madison, Wisconsin 53706} 
\affiliation{Yale University, New Haven, Connecticut 06520} 
\author{T.~Aaltonen}
\affiliation{Division of High Energy Physics, Department of Physics, University of Helsinki and Helsinki Institute of Physics, FIN-00014, Helsinki, Finland}
\author{J.~Adelman}
\affiliation{Enrico Fermi Institute, University of Chicago, Chicago, Illinois 60637}
\author{T.~Akimoto}
\affiliation{University of Tsukuba, Tsukuba, Ibaraki 305, Japan}
\author{B.~\'{A}lvarez~Gonz\'{a}lez$^t$}
\affiliation{Instituto de Fisica de Cantabria, CSIC-University of Cantabria, 39005 Santander, Spain}
\author{S.~Amerio$^z$}
\affiliation{Istituto Nazionale di Fisica Nucleare, Sezione di Padova-Trento, $^z$University of Padova, I-35131 Padova, Italy} 

\author{D.~Amidei}
\affiliation{University of Michigan, Ann Arbor, Michigan 48109}
\author{A.~Anastassov}
\affiliation{Northwestern University, Evanston, Illinois  60208}
\author{A.~Annovi}
\affiliation{Laboratori Nazionali di Frascati, Istituto Nazionale di Fisica Nucleare, I-00044 Frascati, Italy}
\author{J.~Antos}
\affiliation{Comenius University, 842 48 Bratislava, Slovakia; Institute of Experimental Physics, 040 01 Kosice, Slovakia}
\author{G.~Apollinari}
\affiliation{Fermi National Accelerator Laboratory, Batavia, Illinois 60510}
% ...

\collaboration{CDF Collaboration\footnote{With visitors from $^a$University of Massachusetts Amherst, Amherst, Massachusetts 01003,
$^b$Universiteit Antwerpen, B-2610 Antwerp, Belgium, 
$^c$University of Bristol, Bristol BS8 1TL, United Kingdom,
$^d$Chinese Academy of Sciences, Beijing 100864, China, 
$^e$Istituto Nazionale di Fisica Nucleare, Sezione di Cagliari, 09042 Monserrato (Cagliari), Italy,
$^f$University of California Irvine, Irvine, CA  92697, 
$^g$University of California Santa Cruz, Santa Cruz, CA  95064, 
$^h$Cornell University, Ithaca, NY  14853, 
$^i$University of Cyprus, Nicosia CY-1678, Cyprus, 
$^j$University College Dublin, Dublin 4, Ireland,
$^k$University of Edinburgh, Edinburgh EH9 3JZ, United Kingdom, 
$^l$University of Fukui, Fukui City, Fukui Prefecture, Japan 910-0017
$^m$Kinki University, Higashi-Osaka City, Japan 577-8502
$^n$Universidad Iberoamericana, Mexico D.F., Mexico,
$^o$University of Iowa, Iowa City, IA  52242,
$^p$Queen Mary, University of London, London, E1 4NS, England,
$^q$University of Manchester, Manchester M13 9PL, England, 
$^r$Nagasaki Institute of Applied Science, Nagasaki, Japan, 
$^s$University of Notre Dame, Notre Dame, IN 46556,
$^t$University de Oviedo, E-33007 Oviedo, Spain, 
$^u$Texas Tech University, Lubbock, TX  79609, 
$^v$IFIC(CSIC-Universitat de Valencia), 46071 Valencia, Spain,
$^w$University of Virginia, Charlottesville, VA  22904,
$^x$Bergische Universit\"at Wuppertal, 42097 Wuppertal, Germany,
$^{ff}$On leave from J.~Stefan Institute, Ljubljana, Slovenia, 
}}
\noaffiliation

\begin{abstract}
We present a search for associated production of the standard model (SM) Higgs boson and a $Z$ boson where the $Z$ boson decays to two leptons and the Higgs decays to a pair of $b$ quarks in  $p\bar{p}$ collisions at the Fermilab Tevatron.
 We use event probabilities based on SM matrix elements to construct a likelihood function of the Higgs content of the data sample. In a CDF data sample corresponding to an integrated luminosity of 2.7 fb$^{-1}$  we see no evidence  of a Higgs boson with a mass between 100~GeV$/c^2$ and 150~GeV$/c^2$. We set 95\% confidence level (C.L.) upper limits on the cross-section for $ZH$ production as a function of the Higgs
 boson mass $m_H$; the limit is 8.2 times the SM prediction at $m_H = 115$ GeV$/c^2$.
\end{abstract}


\maketitle 

\date{\today}

In the standard model (SM), the Higgs mechanism is responsible for the observed breaking of the $SU(2)_L \otimes U(1)$ symmetry~\cite{PhysRevLett.13.508,PhysRevLett.13.585}, yet the Higgs boson remains the only SM particle that has not been directly observed. Direct searches have set a lower limit on the SM Higgs boson mass $m_H$ of 114.4~GeV$/c^2$ at 95\% C.L.~\cite{lephiggs}, while precision electroweak measurements indirectly constrain its mass to $m_H=76^{+33}_{-24}$~GeV$/c^2$~\cite{indirectlimit}. 
  At hadron colliders the dominant production process for the SM Higgs boson is $gg \rightarrow H$ while its decays are dominated by  $H\to b\bar{b}$ for $m_H<140$~GeV$/c^2$. However, the process $gg \rightarrow H \rightarrow b\bar{b}$ is dwarfed by multi-jet background, necessitating the search for Higgs bosons produced in association with a $W$ or $Z$ boson that decays leptonically. This article reports a search for the process $p\bar{p}\to ZH\to \ell^-\ell^+b\bar{b}$  ($\ell=e,\mu$) in  data with an integrated luminosity of 2.7~fb$^{-1}$ collected with the CDF II detector, nearly 3 times that of the previously reported analysis~\cite{aaltonen:251803}. 
The study of Higgs boson production in association with a $W/Z$ gauge boson for low Higgs boson masses
 is further motivated by the fact that the signal to background ratio is more favorable at the
 Tevatron compared to the Large Hadron Collider.  

 For the first time in a $ZH\to \ell^-\ell^+b\bar{b}$ search, we utilize a method based on leading-order matrix element calculations~\cite{mcfm,madgraph,meref1} convoluted with detector resolution functions~\cite{cdfsim} that form per-event likelihoods. This method, pioneered for use in top quark mass measurements~\cite{Abulencia:2005uq,d0me}, has been recently used in Higgs boson searches in other decay channels~\cite{aaltonen:021802} by forming a discriminating per-event variable.    We extend the technique by expressing the event likelihoods as a function of the $ZH$ signal fraction and maximizing the joint likelihood for the data sample with respect to the signal fraction.  

The CDF II detector~\cite{Acosta:2004yw,coordinates} is an azimuthally and forward-backward symmetric apparatus designed to study $p\overline{p}$ collisions at the Fermilab Tevatron. It consists of a magnetic spectrometer surrounded by calorimeters and muon chambers. The charged particle tracking system, consisting of a silicon detector and drift chamber, is immersed in a 1.4~T magnetic field parallel to the $p$ and $\overline{p}$ beams. Calorimeters segmented in $\eta$ and $\phi$ surround the tracking system and measure the energy of particles detected within them.  The electromagnetic and hadronic calorimeters are lead-scintillator and iron-scintillator sampling devices, respectively. Drift chambers located outside the central hadron calorimeters detect muons. The data used in this analysis are collected with an online selection that requires events to have a lepton with $E_T>18$~GeV (for an electron) or $p_T>18$~GeV$/c$ (for a muon)~\cite{coordinates}.

The event selection used in this analysis closely follows that in Ref.~\cite{aaltonen:251803}. Candidate events are required to have a pair of oppositely charged electrons or muons with invariant mass $76 < m_{\ell \ell} < 106$ GeV$/c^2$. Candidate events are also required to have one jet with $E_T>25$~GeV and at least one additional jet with $E_T>15$~GeV, both within $|\eta|<2.0$. All jet energies are corrected for non-uniformities in calorimeter response, effects from multiple $p\bar{p}$ interactions and for the hadronic energy scale of the calorimeter~\cite{jes}. Candidate events are required to have at least one jet with an associated displaced secondary vertex~\cite{acosta:052003} (``$b$-tags'',  reconstructed using tracks with hits in the silicon detector), thus enriching the $b$-quark content of the sample.

The backgrounds for this analysis are dominated by events with real $Z$ bosons with additional contributions from $t\bar{t}$ and events where an object, such as a jet, is mis-identified as a lepton. We model the backgrounds with events generated with leading-order event generators, normalized to next-to-leading order cross-sections and simulated with a {\sc geant}-based description of the CDF II detector~\cite{cdfsim}.   $Z$+light-flavor jet contributions are modeled with the {\sc alpgen}~\cite{alpgen} simulation code matched with {\sc pythia} in the MLM scheme~\cite{alpgen} for the hadronization and fragmentation.  Heavy flavor contributions from $Z+b\bar{b}$ and $Z+c\bar{c}$ are modeled separately with {\sc alpgen} and combined with the light-flavor jet samples.  The $WZ$, $ZZ$ and $t\bar{t}$ processes are modeled using {\sc pythia}~\cite{pythia}. Events where a jet is mis-identified as a charged lepton are modeled using jet-enriched data samples~\cite{abulencia:032008,aaltonen:251803}. We model the kinematics of $ZH\to \ell^+\ell^-b\bar{b}$ events using {\sc pythia} for $m_H$ ranging from 100~GeV$/c^2$ to 150~GeV$/c^2$. The  signal and background contributions expected in 2.7~fb$^{-1}$ and the number of observed events are given  in Table~\ref{tab:sample}.

\begin{table}
\caption{Expected and observed numbers of events with 1 or 2 $b$-tagged jets in 2.7 fb$^{-1}$ of data. The $ZH$ expectation is shown for $m_H=115$~GeV$/c^2$ assuming the production cross section at $\sqrt{s}=1.96$~TeV for $q\bar{q}\to Z^{*}\to ZH$ to be 1.04~pb~\cite{higgsxsec} and the branching ratio $\mathcal{B}(H\to b\bar{b})$ to be 73\%~\cite{Djouadi199856}.}
\begin{ruledtabular}
\begin{tabular}{lcc}
Source & 1 tag & $\geq 2$ tag \\
\hline
$Z\to\ell^+\ell^-+$light partons & $129.6\pm 24.0$ & $5.5 \pm 0.9$\\
$Z\to\ell^+\ell^-+b\bar{b},c\bar{c}$ & $107.2\pm 14.0$ & $19.5\pm 3.4$\\
$ZZ,WZ$ &  $11.6 \pm 1.3$  &  $2.9 \pm 0.4$ \\
$t\bar{t}$ &  $13.9 \pm 2.0$  & $7.7 \pm 1.1$  \\
Mis-ID lepton & $15.9\pm 6.5$ & $0.4\pm 0.2$ \\
$ZH$ & $1.3 \pm 0.2$ & $0.7 \pm 0.1$ \\
\hline
Total expected & $279.5 \pm 28.6 $ &$36.3 \pm 3.7 $ \\
Data & 258 & 32 
\end{tabular}
\end{ruledtabular}
\label{tab:sample}
\end{table}%

We denote the $ZH$ signal probability by $P_{ZH}({\bf x_i}|m_H)$ where $m_H$ is a parameter and ${\bf x_i}$ 
 represents  the collection of the measured 4-vector momenta of the two selected leptons, the two selected jets,
and the two components of the missing transverse momentum, in a given event $i$.
Similarly we denote the background probability as $P_b(\mathbf{x_i})$.  The per-event likelihood as a function 
 of the signal fraction $s$ for a given event $i$ is

\begin{equation}
L(s, {\bf x_i}| m_H) = s P_{ZH}({\bf x_i}|m_H) +  (1-s) P_{b}({\bf x_i}).
\label{eventP}
\end{equation}
\noindent We evaluate $P_{ZH}$ and $P_b$ by convoluting the leading-order matrix elements for the process with detector resolution functions and integrating over unmeasured quantities. Thus, $P_{ZH}$ is a probability density in ${\bf x_i}$ and can be expressed as 
\begin{equation}
\begin{split}
P_{ZH}(\mathbf{x_i}|m_H)=&\frac{1}{\sigma(m_H)} \int d\Phi |\mathcal{M}_{ZH}(q, p; m_H)|^2 \\
&\times \prod_j [W(p_j,\mathbf{x_i})] f_{PDF}(q_1)f_{PDF}(q_2)
\end{split}
\label{matrixelement}
\end{equation}
\noindent where $\mathcal{M}_{ZH}$ is the leading-order matrix element for the process $q\bar{q}\to ZH\to \ell^+\ell^-b\bar{b}$ evaluated for a pair of incoming partons $q$ and outgoing particles $p$, $W(p_j, \mathbf{x_i})$ are transfer functions~\cite{Abulencia:2006qj} linking the outgoing particle momenta $p_j$ to measured quantities $\mathbf{x_i}$ and the $f_{PDF}$ are parton density functions of the incoming partons. The factor $1/\sigma(m_H)$ ensures that the probability density satisfies the normalization condition, $\int d\mathbf{x_i}_i\ P_{ZH}(\mathbf{x_i}|m_H) = 1$.

 The sample likelihood $\mathcal{L}$ is obtained by taking the product over all events $i$ in the sample 

\begin{equation}
{\cal L} (s| m_H) = \prod_{i} L(s,{\bf x_i}| m_H).
\label{sampleP}
\end{equation}

\begin{figure}
\begin{center}
\hskip -0.0cm
\begin{minipage}{0.48\columnwidth}
\includegraphics[width=1.05\columnwidth]{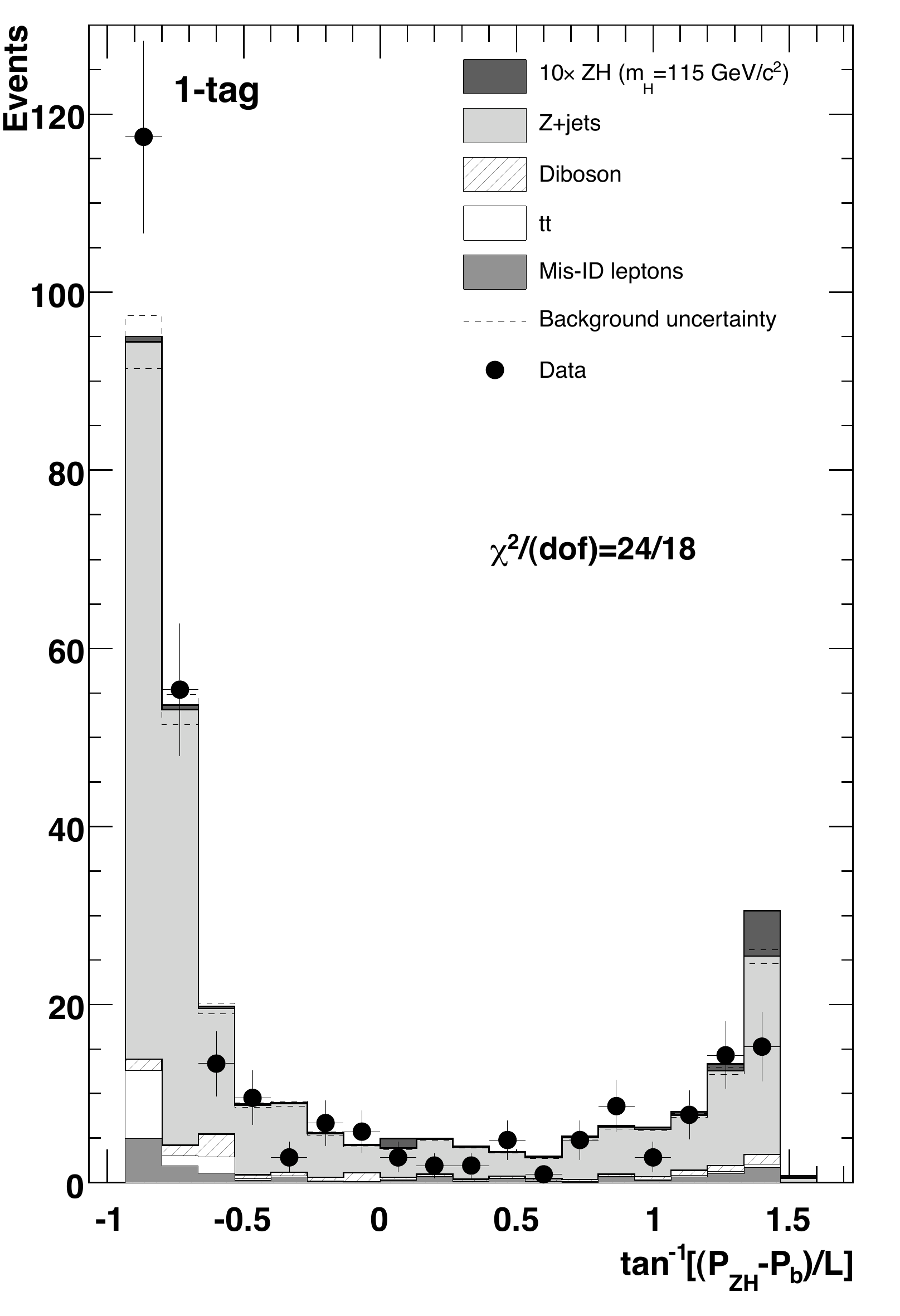}
\end{minipage}
\begin{minipage}{0.48\columnwidth}
\includegraphics[width=1.05\columnwidth]{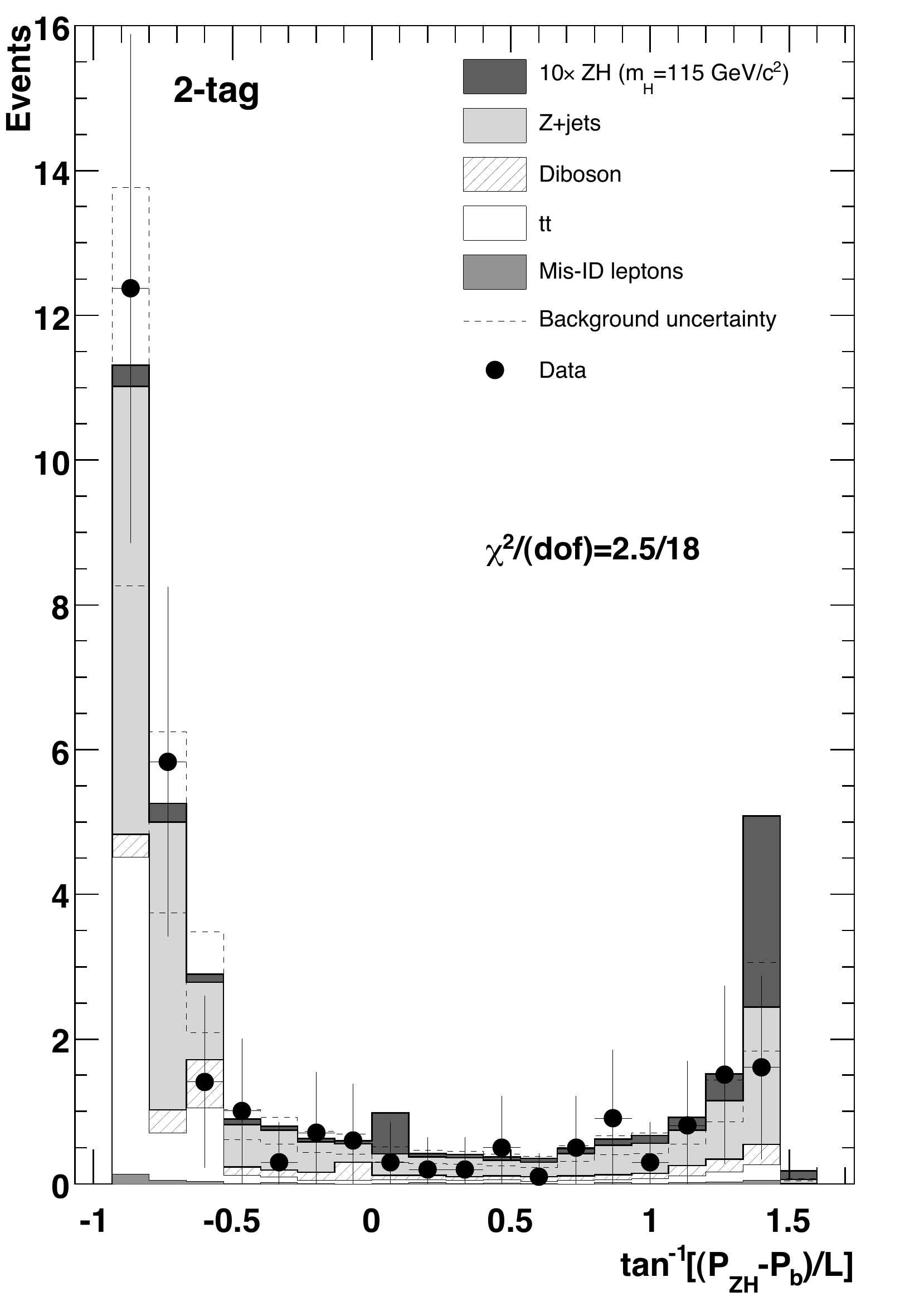}
\end{minipage}
\caption{The distribution of $\tan^{-1}\cal D$ where the discriminant $\cal D$~$=(P_{ZH}-P_b)/L$ for expected backgrounds and data for events with one (left)  and two (right) $b$-tags. The expected signal ($\times 10$) is overlaid.}
\label{lhood}
\end{center}
\end{figure}

\noindent  We enhance our statistical sensitivity by exploiting the expected difference in the rate of signal and background events with two $b$-tagged jets. We replace $P_{ZH}({\bf x_i}|m_H)$ by $P_{ZH}({\bf x_i}, n |m_H) 
 \equiv P_{ZH}({\bf x_i}|m_H) \cdot P_{ZH}( n|m_H)$
 and $P_{b}({\bf x_i})$ by $P_{b}({\bf x_i}, n) \equiv P_{b}({\bf x_i}) \cdot P_{b}(n)$, where
 $P_{ZH}(n|m_H)$ [$P_{b}(n)$]
 denotes the probability of tagging signal [background] events with $n$ tags. Table~\ref{tab:tagrates} shows the expected tagging rates for simulated signal and background event samples. 
\begin{table}
\caption{Expected single- and double-tag probabilities,  $P(n=1)$ and $P(n \geq 2)$,  for signal and background events passing our selection. $Z\to\ell^+\ell^-+$jets includes jets from both light and heavy quarks.}
\begin{ruledtabular}
\begin{tabular}{lcc}
Source & $P(n=1)$ & $P(n \geq 2)$ \\
\hline
$Z\to\ell^+\ell^-+$jets & $0.91$ & $0.09$\\
$WZ,ZZ$ &  $0.80$  &  $0.20$ \\
$t\bar{t}$ &  $0.74$  & $0.26$  \\
$ZH~(m_H = 100$~GeV$/c^2)$ & $0.67$ & $0.33$ \\
$ZH~(m_H = 125$~GeV$/c^2)$ & $0.65$ & $0.35$ \\
$ZH~(m_H = 150$~GeV$/c^2)$ & $0.63$ & $0.37$ \\
\end{tabular}
\end{ruledtabular}
\label{tab:tagrates}
\end{table}

 The measured signal fraction $S_{meas}$ is the value of $s$ which maximizes ${\cal L}$$(s| m_H)$. Using Eq.~(\ref{eventP}), we can define a per-event discriminant $\cal D$$_i$$\equiv \partial \ln L / \partial s = (P_{ZH} - P_b) / L$ which increases (decreases) for more signal-like (background-like) events. The maximum-likelihood estimator for the measured signal fraction $S_{meas}$ corresponds to 
 $\Sigma_i$$\cal D$$_i |_{s=S_{meas}}=0$. The distribution of $\tan^{-1}\cal D$$_i (s = S_{meas})$ for simulated events and data is shown in Fig.~\ref{lhood}.

The dominant backgrounds in our data sample are due to $Z+$jets, $t \bar{t}$ and $ZZ$ processes, in the expected proportions denoted by $\lambda_{Zjj}$, $\lambda_{t \bar{t}}$ and $\lambda_{ZZ}$ respectively. The background probability in Eq.~(\ref{eventP}) is given by

\begin{equation}
P_{b}({\bf x_i}, n) = \lambda_{Zjj} P_{Zjj}({\bf x_i}, n) +  \lambda_{t\bar{t}} P_{t\bar{t}}({\bf x_i}, n) + \lambda_{ZZ} P_{ZZ}({\bf x_i},n ),
\label{backgroundP}
\end{equation}

\noindent where  $P_{Zjj}({\bf x_i}, n)$, $P_{t \bar{t}}({\bf x_i}, n)$ and $P_{ZZ}({\bf x_i}, n)$ are the respective probability densities (normalized to unit integral) for the $Z+$jets, $t \bar{t}$, and $ZZ$ background processes with $n$ tags. 
 Normalization of $P_{b}$ is ensured by requiring $\lambda_{Zjj} + \lambda_{t \bar{t}} + \lambda_{ZZ} = 1$. 

We construct confidence intervals~\cite{PhysRevD.57.3873} for the test statistic $R=\mathcal{L}(S_{meas}|S_{true})/\mathcal{L}(S_{meas}|S_{true}^{best})$  by performing simulated experiments with the expected proportions of background and varying the amounts of signal, such that $S_{true}$ is the true (input) signal fraction in the simulated experiment. 
 $S_{true}^{best}$ is the input signal fraction that has the highest likelihood for a given measured signal fraction, $S_{meas}$. $\mathcal{L}(S_{meas}|S_{true})$ is given by Eq.~(\ref{sampleP}) for the simulated experiment with the chosen value of $S_{true}$ and $m_H$. 
 Since we are measuring the fractional signal content in the data sample, 
 the number of events in each simulated experiment is held fixed at the value of 290 events observed
 in the data. 

\begin{figure}
\begin{center}
\includegraphics[width=1.0\columnwidth]{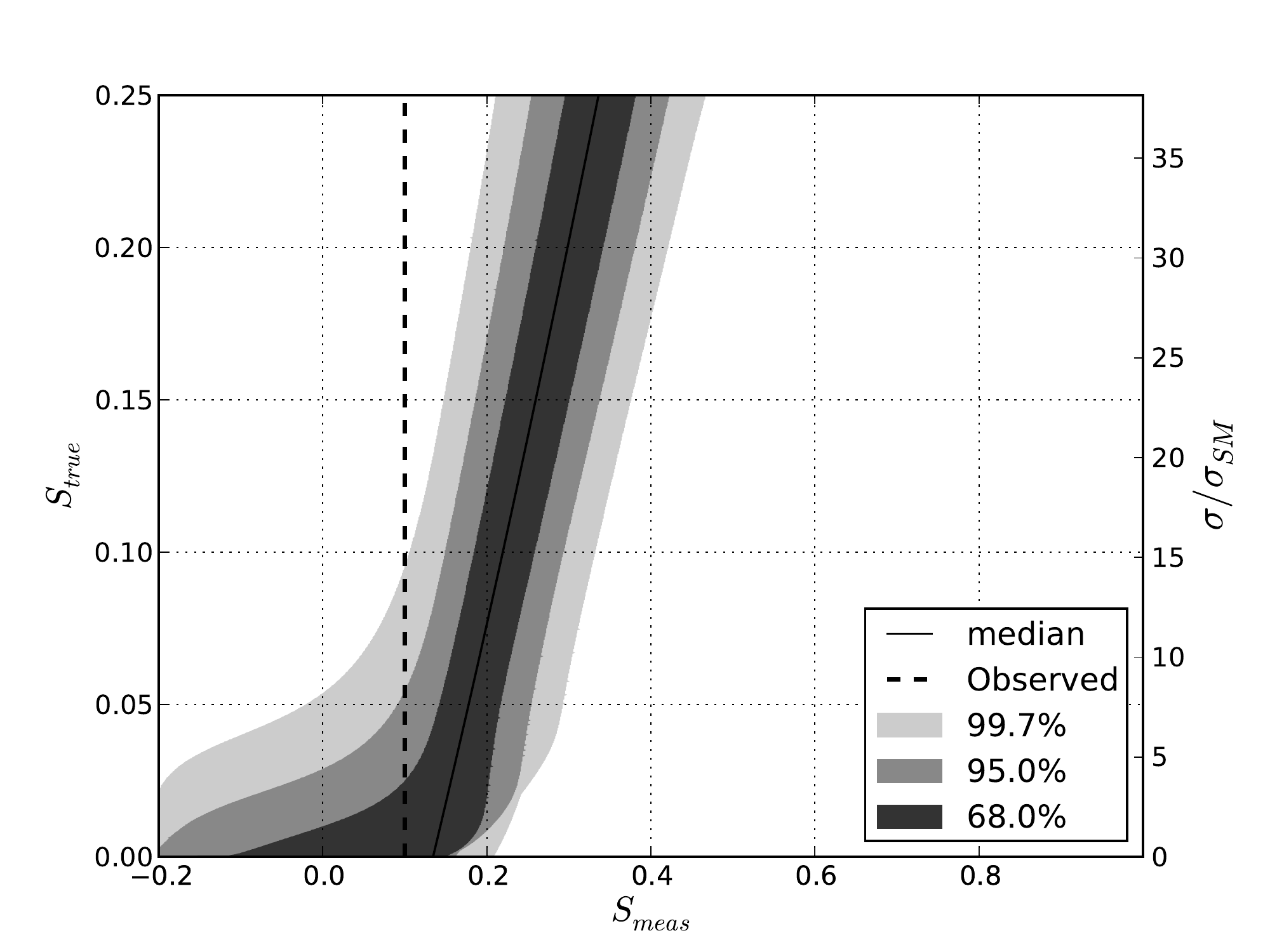}
\caption{Confidence intervals in the measured signal fraction $S_{meas}$ (along the $x-$axis) 
 with 68\% C.L., 95\% C.L. and 99.7\% C.L., for a range of true signal fraction $S_{true}$ values (along the $y-$axis
 on the left)
 chosen in the simulated experiments. The signal cross section ratio equivalent to $S_{true}$ is shown on the $y-$axis on
 the right. The intervals shown here are computed for a Higgs boson mass 
 of $m_H=115$~GeV$/c^2$, and include statistical and systematic 
 uncertainties. The vertical dashed line indicates the value of $S_{meas}$ obtained from the data.  
 }
\label{fc115}
\end{center}
\end{figure}

 The methodology from Ref.~\cite{PhysRevD.57.3873} is used to construct confidence intervals in $S_{meas}$ for each
 chosen value of $S_{true}$ and $m_H$. This method removes any bias resulting from imperfections in our modeling
 by relating $S_{meas}$ to $S_{true}$. The confidence intervals in $S_{meas}$ obtained for $m_H=115$~GeV$/c^2$ 
 and $0 \le S_{true} \le 0.25$ are shown in Fig.~\ref{fc115}. For a given value of $S_{meas}$ obtained from the data (or
 from an independent simulated experiment to evaluate the {\it a priori} expectation), we extract 
 the range of $S_{true}$ for which the confidence intervals contain this value of $S_{meas}$. A feature of this method
 is that the resulting range of $S_{true}$ can be quoted as an upper limit on $S_{true}$ (if the lower bound is zero)
 or as a two-sided measurement of $S_{true}$. As Fig.~\ref{fc115} shows, we obtain an upper limit on $S_{true}$ given the data,
 which we convert to the equivalent upper limit on the signal cross section. This procedure is repeated for 
 the range of Higgs boson masses $100 \le m_H \le 150$~GeV$/c^2$. 

We evaluate systematic uncertainties by varying process rates and kinematic distributions in our
 simulated experiments. We apply a rate uncertainty of 40\% for $Z$ boson events and of 20\% for diboson and
 $t\bar{t}$ events. The
 uncertainty on the rate of heavy flavor production in association with a gauge boson is
 based on comparisons of data with theoretical predictions~\cite{abulencia:032008}. The uncertainty on the diboson
 and $t\bar{t}$ contribution includes the uncertainties in the cross sections, selection efficiencies and 
 the top quark mass~\cite{aaltonen:251803}. 
A rate uncertainty of 50\% is applied for mis-identified lepton events due to the uncertainty on the 
 lepton misidentification probability~\cite{aaltonen:251803}. A rate uncertainty of 6\% due to the luminosity uncertainty is applied to all events. The per-jet uncertainty on the $b$-tagging efficiency is 8\% for events  with $b$ partons, 16\% for events with $c$ partons and 13\% for events with no heavy flavor~\cite{aaltonen:251803}.  Our analysis is weakly sensitive to uncertainties in the expected total number of events passing our selection, since it relies only on the shapes of measured distributions. 
 Uncertainties in the shapes of kinematic distributions are propagated by varying the amount of QCD radiation in simulated signal events and the jet energy scale in simulated signal and background events within their respective
  uncertainties~\cite{jes}.

We evaluate confidence intervals for a range of Higgs masses between 100~GeV$/c^2$ and 150~GeV$/c^2$. We evaluate 
 {\it a priori} 95\% C.L. upper limits on the cross section for the process $p \bar{p} \to ZH\to \ell^+\ell^-b\bar{b}$.
We express these limits as a ratio with respect to the SM prediction.  
 These expected limits along with those observed in the data are shown in Table~\ref{limits}.

\begin{table}
\caption{Upper limits at 95\% C.L. on the $ZH\to \ell^+\ell^-b\bar{b}$ cross section, 
 shown as a ratio to the SM cross section. The column labelled ``Expected'' shows the median of the limits
 obtained from simulated experiments containing no signal, and the columns labelled ``$\pm 1 \sigma$'' show the range containing 68\%
 of the expected limits. }
\begin{ruledtabular}
\begin{tabular}{ccccc}
$m_H$ [GeV$/c^2$]  & $-1 \sigma$      & Expected               & $+1 \sigma$            &      Observed  \\
             & $[\sigma/\sigma_{SM}]$ & $[\sigma/\sigma_{SM}]$ & $[\sigma/\sigma_{SM}]$ & $[\sigma/\sigma_{SM}]$ \\
\hline
100 & 6.0 & 8.7  & 12.4 & 7.0\\
105 & 6.0 & 8.7  & 12.9 & 6.5\\
110 & 7.5 & 11.3 & 16.8 & 7.6 \\
115 & 8.3 & 12.1 & 18.2 & 8.2\\
120 & 9.3 & 13.5 & 20.0 & 9.0\\
125 & 13.2 & 18.3 & 27.1 & 13.2\\
130 & 17.1 & 24.2 & 35.7 & 17.7\\
135 & 21.8 & 31.0 & 44.8 & 22.9\\
140 & 31.0 & 44.3 & 65.4 & 32.0\\
145 & 42.8 & 61.6 & 89.9 & 43.1\\
150 & 73.7 & 104 & 153 & 71.3\\
\end{tabular}
\end{ruledtabular}
\label{limits}
\end{table}

In conclusion, we have performed a search for the SM Higgs boson decaying to $b\bar{b}$ produced in association with a $Z$ boson. This is the first analysis performed in this channel with a matrix element method. The data  show no excess over expected non-Higgs backgrounds. We set  95\% C.L. upper limits on the cross section of this process for a range of Higgs boson masses.  The limit at $m_H = 115$~GeV$/c^2$ is 8.2 times greater than the SM prediction. This result improves by a factor of 2 over the previously published result in this channel~\cite{aaltonen:251803}.
We are exploring further improvements in this technique by separating the leading-order and next-to-leading order
contributions to the signal and backgrounds, as well as the use of matrix-element-based probabilities in conjunction
with other multivariate discriminants.

We thank the Fermilab staff and the technical staffs of the participating institutions for their vital contributions. This work was supported by the U.S. Department of Energy and National Science Foundation; the Italian Istituto Nazionale di Fisica Nucleare; the Ministry of Education, Culture, Sports, Science and Technology of Japan; the Natural Sciences and Engineering Research Council of Canada; the National Science Council of the Republic of China; the Swiss National Science Foundation; the A.P. Sloan Foundation; the Bundesministerium f\"ur Bildung und Forschung, Germany; the World Class University Program, the National Research Foundation of Korea; the Science and Technology Facilities Council and the Royal Society, UK; the Institut National de Physique Nucleaire et Physique des Particules/CNRS; the Russian Foundation for Basic Research; the Ministerio de Ciencia e Innovaci\'{o}n, and Programa Consolider-Ingenio 2010, Spain; the Slovak R\&D Agency; and the Academy of Finland.

\bibliography{prd_zhme}

\begin{thebibliography}{23}
\bibitem[{coo()}]{coordinates}
\bibinfo{note}{CDF uses a cylindrical coordinate system with the $z$ axis along
  the proton beam axis. Pseudorapidity is $\eta \equiv -\ln(\tan \theta/2)$,
  where $\theta$ is the polar angle, and $\phi$ is the azimuthal angle relative
  to the proton beam direction, while $p_T = |p|\ \sin \theta$, $E_T = E\ \sin
  \theta$.}
\end{thebibliography}
\end{document}